\documentclass[aps,prl,twocolumn,groupedaddress]{revtex4-1}
\usepackage{amsmath} 
\usepackage{graphicx}
\usepackage{dcolumn}
\usepackage{mathrsfs}
\usepackage{bm}
\usepackage{subfigure}
\usepackage{natbib}
\usepackage{chemfig}
\usepackage[colorlinks,linkcolor=blue,anchorcolor=blue,citecolor=blue,urlcolor=blue]{hyperref}

\begin{document}
\title{Stationary entanglement between light and microwave via ferromagnetic magnons}
\author{Qizhi Cai$^{1}$}
\author{Jinkun Liao$^{1}$}\thanks{E-mail: jkliao@uestc.edu.cn}
\author{Qiang Zhou$^{1,2,}$}\thanks{E-mail: zhouqiang@uestc.edu.cn}
\address{$^{1}$School of optoelectronic science and engineering, University of Electronic Science and Technology of China, Chengdu, Sichuan, China}
\address{$^{2}$Institute of Fundamental and Frontier Sciences, University of Electronic Science and Technology of China, Chengdu, Sichuan, China}

\date{\today}
\begin{abstract}
We show how to generate stationary entanglement between light and microwave in a hybrid opto-electro-magnonical system which mainly consists of a microwave cavity, a yttrium iron garnet (YIG) sphere and a nanofiber. The optical modes in nanofiber can evanescently coupled to whispering gallery modes, that are able to interact with magnon mode via spin-orbit interaction, in YIG sphere, while the microwave cavity photons and magnons are coupled through magnetic dipole interaction simultaneously. Under reasonable parameter regimes, pretty amount of entanglement can be generated, and it also shows persistence against temperature. Our work is expected to provide a new perspective for building more advanced and comprehensive quantum networks along with magnons for fast-developing quantum technology and for studying the macroscopic quantum phenomena.
\end{abstract}

\pacs{Valid PACS appear here}
\maketitle

\section{Introduction}
The entanglement between light and microwave attracts considerable research interests, mainly owing to its indispensability not only for investigating the fundamental physics, but also for the fast-developing quantum technologies. For example, the optical photons in fiber can connect various distant solid-state qubits that typically operate at microwave band, such as superconducting circuits, NV centers and quantum dots, which permit to efficiently perform the gates and operations used in quantum information processing. Consequently, the light-microwave (L-M) entanglement is of vital importance to build scalable and extensive hybrid quantum systems \citep{ref1} for realizing the quantum computer \citep{ref2, ref3}, quantum communication \citep{ref4} and quantum internet \citep{ref5, ref6, ref7, ref8}. On the other hand, this kind of entanglement has the potential to enhance the performance of detecting weak radio-frequency signals \citep{ref9, ref10}, illuminating the far-reaching targets \citep{ref11, ref12, ref13, ref14} and improving the non-invasive diagnostic scanner in biomedical applications \citep{ref15}. However, due to the five-order energy gap, the entanglement between the optical and microwave modes can not be generated directly, one often needs the intermediate systems, which well couples to both of them, to complete this process. Over the decades, various theoretical proposals and experimental realizations for this kind of intermediate systems have arised, including the optomechanical systems \citep{ref16, ref17, ref18, ref19, ref20, ref21} and the electro-optical systems \citep{ref22, ref23}.

Recently, a new series of hybrid quantum systems based on ferromagnetic crystals, particularly the yttrium iron garnet (YIG) sphere, have emerged as promising candidates for novel quantum technologies \citep{ref24}. The quanta of collective spin excitations in YIG sphere, called magnons, can coherently couple to the microwave photons inside microwave cavities and reach the strong-coupling regime \citep{ref25, ref26, ref27, ref28}, which can be utilized to realize the entanglement between microwave photons and magnons \citep{ref29}. These features manifest the potential of such systems, also named cavity electromagnonical systems, for quantum technology platforms, somehow similar with the electromechanical systems \citep{ref30}. Furthermore, due to the resemblance with optomechanical systems \citep{ref31}, the optomagnonical systems are high investigated both from theoretical and experimental perspective in recent years, where the optical photons and magnons couple to each other via Faraday effect and, the same with electromagnonical counterpart, their interaction can be enhanced by optical cavities \citep{ref32, ref33, ref34, ref35, ref36, ref37, ref38}. We notice that, on the basis of the electro- and optomagnonical couplings mentioned above, Nakamura and his co-workers realized the bidirectional conversion between microwave and light via ferromagnetic magnons \citep{ref39}, which inspires us to wonder whether we can generate entanglement between light and microwave by means of the magnons.

Here, we propose a scheme to generate the entanglement between light and microwave based on a hybrid system composed of electro- and optomagnonical systems. We utilize the quantum Langevin equations to describe the dynamics of this system, then solve the linearized dynamics and obtain the L-M entanglement in a steady state. This work is expected to provide a new perspective to entangle electromagnetic modes in microwave and optical frequency domain, which is essential to promote the development of quantum technology as discussed above.

This paper is organized as follows. Sec. II shows the physical model and simple sketch of our hybrid system and give the quantum Langevin equations describing the dynamics of the system. In Sec. III, we derive the covariance matrix to obtain the logarithmic negativity of interested bipartite subsystems, which is regarded as the entanglement measure in this work. In Sec. IV, we analyze the impact of key parameters on the entanglement and briefly discuss the detection of it, while Sec. V is for conclusion.

\section{Model}
\begin{figure}
	\centering
	\includegraphics[width=1\linewidth]{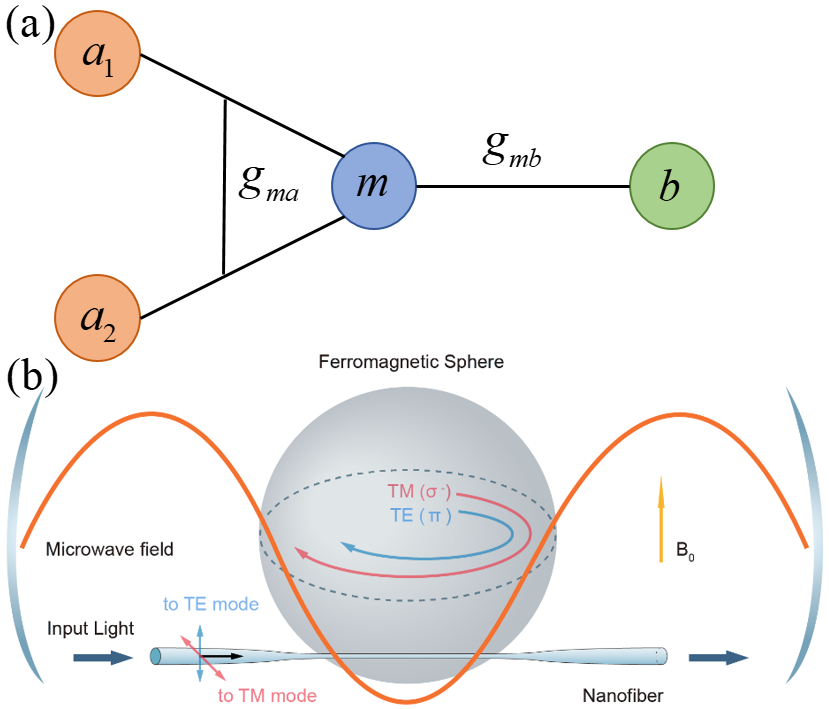}
	\caption{(a) The schematic of the light-magnon-microwave interface, where $a_1$ and $a_2$ represent two optical modes, TE and TM modes, $m$ indicates the magnon mode and $b$ means the microwave mode. Two optical modes $a_j$ ($j$=1,2) are coupled to a magnon mode $m$ via the three-wave process with couplings $g_{ma}$ depicted in the triangle, while the microwave mode $b$ electromagnonically coupled to $m$ with coupling rate $g_{mb}$, simultaneously. (b) Sketch of the system. A ferromagnetic sphere, YIG sphere in this work, is placed in a microwave cavity, where the magnon-microwave coupling is established by a external bias magnetic field $B_0$. The optical modes in nanofiber are evanescently coupled to TE and TM whispering gallery modes (WGMs) in YIG sphere, in which two WGMs share a three-wave process with the magnon mode.}
	\label{fig:Fig1.png}
\end{figure}
A schematic of our hybrid magnon-based system is shown in Fig. 1(a). In the microwave cavity, microwave mode $b$ is electromagnonically coupled to the magnon mode $m$ inside a YIG sphere, while two optical modes $a_j$ ($j$=1,2) share a three-wave process, the Brillouin scattering process, with magnon mode $m$ at the same time. To be more specific in the optomagnonical part, the situation we consider is that these WGM photons interact with the magnons when the external magnetic field is applied perpendicularly to the WGM orbit plane. In this way, magnons could interact with virtually purely $\sigma^{+}$-, $\sigma^{-}$- or $\pi$-polarized photon depending on the polarization of photons and the direction of the WGM orbit. The direction of WGM orbits has two: counterclockwise (CCW) orbit corresponding to (TE,TM) = ($\pi$, $\sigma^{+}$) resonant in the YIG sphere and clockwise (CW) orbit corresponding to (TE,TM) = ($\pi$, $\sigma^{-}$). In the CCW case, the interaction Hamiltonian reads $H^{CCW}_{int} = \hbar g_{ma} (\hat{a}^{\dagger}_{TM} {\hat{a}^{ }}_{TE} \hat{m} + {\hat{a}^{ }}_{TM} \hat{a}^{\dagger}_{TE} \hat{m}^{\dagger})$ with coupling constant $g_{ma}$, which means that if the input photon is in the TM mode with angular frequency $\omega$ that is $\sigma^{+}$-polarized in the resonator, via the Brillouin scattering, one magnon with angular frequency $\omega_{mag}$ and one down-converted photon with $\pi$-polarization in the TE mode and with angular frequency $\omega-\omega_{mag}$ are generated, fulfilling the conservation of energy and spin angular momentum. In the CW case that is symmetrial with the CCW one, the interaction Hamiltonian is $H^{CW}_{int} = \hbar g_{ma} (\hat{a}^{\dagger}_{TE} {\hat{a}^{ }}_{TM} \hat{m} + {\hat{a}^{ }}_{TE} \hat{a}^{\dagger}_{TM} \hat{m}^{\dagger})$, and the interaction read that if the input photon is in the TE mode that is $\pi$-polarized in the resonator, via the Brillouin scattering, one magnon and one down-converted photon with $\sigma^{-}$-polarization in the TM mode are generated \citep{ref36}.

For simplicity and without loss of generality, in this work, we only choose the CW case for generating the L-M entanglement as shown in the Fig.1(b), which means we would only pump the light that couples the TE WGM in the YIG sphere. The total Hamiltonian of the system is $H = H_{0} + H_{int}$, where the free energy Hamiltonian is
\begin{equation}
\begin{split}
H_{0}=\sum\nolimits_{j}\hbar \omega _{aj} \hat{a}^{\dagger}_{j}\hat{a}_j+\hbar \omega _{m} \hat{m}^{\dagger}\hat{m}+\hbar \omega _{b} \hat{b}^{\dagger}\hat{b}, 
\end{split}
\end{equation}
and the interaction Hamiltonian read \citep{ref29, ref36}
\begin{equation}
\begin{split}
H_{int} = \hbar g_{ma}(\hat{a}_1 \hat{a}^{\dagger}_2 \hat{m}^{\dagger} + \hat{a}^{\dagger}_1 \hat{a}_2 \hat{m}) + g_{mb}(\hat{b}+\hat{b}^{\dagger})(\hat{m}+\hat{m}^{\dagger}),
\end{split}
\end{equation}
in which $\hat{a}_1$, $\hat{a}_2$, $\hat{b}$ and $\hat{m}$ ($\hat{a}^{\dagger}_1$, $\hat{a}^{\dagger}_2$, $\hat{b}^{\dagger}$ and $\hat{m}^{\dagger})$ are the annihilation (creation) operators of the TE optical, TM optical, microwave and magnon modes as shown in Fig.1(b), respectively, satifying $[\hat{O}, \hat{O}^{\dagger}]=1$ ($O=a_j,b,m$). $\omega_{aj}$ and $\omega_{b}$ are optical and microwave resonance frequency, and the magnon frequency is determined by the bias magnetic field $B_0$ and gyromagnetic ratio $\gamma / 2 \pi$ = 28 GHz/T, with $\omega_m = \gamma B_0$. The electromagnonical coupling $g_{mb}$ can get larger than the damping rates of microwave mode $\kappa_b$ and of magnon mode $\kappa_m$, which guarantees the strong coupling regime between the microwave and magnon \citep{ref25, ref26, ref27, ref28}. The optomagnonical coupling is described by \citep{ref36}
\begin{equation}
\begin{split}
{g_{ma}} = \mathcal{V}\frac{c}{{{n_r}}}\sqrt {\frac{2}{{{n_{spin}}V_{sp}}}},
\end{split}
\end{equation}
where the YIG's Verdet constant $\mathcal{V}$ = 3.77 rad/cm, the refractive index $n_r$ = 2.19, and the spin density $n_{spin}$ = 2.1$\times {10^{28}}/{m^3}$. $V_{sp}=\frac{4\pi}{3}r^{3}$ is the volume of the YIG sphere, $r$ is the radius of YIG sphere, and $c$ is the speed of light in vacuum.

We can linearize optomagnonical interaction $H_{ma} = \hbar g_{ma}(\hat{a}_1 \hat{a}^{\dagger}_2 \hat{m}^{\dagger} + \hat{a}^{\dagger}_1 \hat{a}_2 \hat{m})$ by assuming the $a_1$ mode is resonantly pumped by a strong optical field, under this condition, $\hat{a}_1$ can be regarded as a complex number $\alpha = \left \langle \hat{a}_1 \right \rangle$. Thus, the optomagnonical interaction can be rewriten as $H^{'}_{ma} = \hbar g_{ma} \alpha (\hat{a}^{\dagger}_2 \hat{m}^{\dagger} + \hat{a}_2 \hat{m})$, where we have chosen the proper phase reference so that $\alpha$ can be taken real and positive. By applying the rotating-wave approximation, the electromagnonical interaction $g_{mb}(\hat{b}+\hat{b}^{\dagger})(\hat{m}+\hat{m}^{\dagger})$ changes to $g_{mb}(\hat{b}\hat{m}^{\dagger}+\hat{b}^{\dagger}\hat{m})$, under the condition that $\omega_b$, $\omega_m$ $\gg$ $g_{mb}$, $\kappa_b$, $\kappa_m$, where $\kappa_b$ and $\kappa_m$ are the damping rates of microwave and magnon mode \citep{ref29}. After these operations, the interaction Hamiltonian become
\begin{equation}
\begin{split}
H^{'}_{int} = \hbar G_{ma} (\hat{a}^{\dagger} \hat{m}^{\dagger} + \hat{a} \hat{m}) + g_{mb}(\hat{b}\hat{m}^{\dagger}+\hat{b}^{\dagger}\hat{m}),
\end{split}
\end{equation}
where we have denoted $\hat{a}_2$ as $\hat{a}$ and $g_{ma} \alpha$ as $G_{ma}$ for simplicity. We notice that the optomagnonical interaction is a two-mode squeezing term that can generate entanglement between $\hat{a}$ and $\hat{m}$, while the electromagnonical interaction is a beam-splitter term that can transfer the non-classical correlation between $\hat{b}$ and $\hat{m}$. The entanglement between $\hat{a}$ and $\hat{m}$, created by two-mode squeezing, can transfer to $\hat{b}$ via the beam-splitter-type coupling between modes $\hat{b}$ and $\hat{m}$, therefore, building the entanglement between light and microwave.

In the interaction picture with respect to $H_{ip} = \hbar \omega_{p} \hat{a}^{\dagger}_1\hat{a}_1 + \hbar \omega _{dm} \hat{m}^{\dagger}\hat{m}+\hbar \omega _{db} \hat{b}^{\dagger}\hat{b}$, the quantum Langevin equations (QLEs) describing the dynamics of the hybrid system are given by
\begin{equation}
\begin{split}
&\dot{\hat{m}} = -(i \Delta_m + \kappa_m) \hat{m} - i G_{ma} \hat{a}^{\dagger} - i g_{mb} \hat{b} + \sqrt {2{\kappa _m}} {m^{in}},\\
&\dot{\hat{a}} = -(i \Delta_a + \kappa_a) \hat{a} - i G_{ma} \hat{m}^{\dagger} + \sqrt {2{\kappa _a}} {a^{in}},\\
&\dot{\hat{b}} = -(i \Delta_b + \kappa_b) \hat{b} - i g_{mb} \hat{m} + \sqrt {2{\kappa _b}} {b^{in}},
\end{split}
\end{equation}
where $\Delta_m = \omega_m - \omega_{dm}$, $\Delta_a = \omega_{a2} - \omega_{da}$ and $\Delta_b = \omega_b - \omega_{db}$ are the detuning of magnon, optical TM and microwave modes, with driving angular frequency $\omega_{dm}$, $\omega_{da}$ and $\omega_{da}$, resectively. $\kappa_a = \omega_{a2} / Q_a$ is the damping rates of optical TM mode, $Q_a$ is the quality factor of YIG sphere for WGM. The input noise terms of each mode are $a^{in}$, $b^{in}$ and $m^{in}$, which can be regarded as zero-Gaussian process, characterized by the following correlation \citep{ref40}
\begin{equation}
\begin{split}
&\left \langle m^{in}(t)m^{in,\dagger}(t')\right \rangle=[N(\omega_m)+1]\delta(t-t'),\\ 
&\left \langle m^{in,\dagger}(t)m^{in}(t')\right \rangle=N(\omega_m)\delta(t-t'),\\
&\left \langle a^{in}(t)a^{in,\dagger}(t')\right \rangle=[N(\omega_a)+1]\delta(t-t'),\\ 
&\left \langle a^{in,\dagger}(t)a^{in}(t')\right \rangle=N(\omega_a)\delta(t-t'),\\
&\left \langle b^{in}(t)b^{in,\dagger}(t')\right \rangle=[N(\omega_{b})+1]\delta(t-t'),\\
&\left \langle b^{in,\dagger}(t)b^{in}(t')\right \rangle=N(\omega_{b})\delta(t-t'),
\end{split}
\end{equation}
in which $N({\omega _k})=1/[$exp$(\hbar\omega_k/k_B T)-1]$ ($k=m,a,b$) are the mean thermal optical photon, microwave photon, and magnon number, respectively. We can safely assume that $N({\omega _a})\simeq 0$ thanks to $\hbar \omega _c/ k_B T \gg 1$, while $N({\omega _b})$ and $N({\omega _m})$ cannot be neglected even the environment temperature $T$ is quite low.

\section{Covariance matrix of the system and quantification of light-microwave entanglement}
By common interaction with magnon mode, the entanglement between light and microwave can be generated, in other word, the quantum correlations among appropriate quadratures of the optical and microwave fields. Based on this, we concentrate our focus on the quadratures of each mode, and introduce the quadrature $\hat X_m = ( \hat{m} + \hat{m}^{\dagger} ) / \sqrt{2}$ and $\hat Y_m = ( \hat{m} - \hat{m}^{\dagger} ) / i \sqrt{2}$ for the magnon mode, $\hat X_a = ( \hat{a} + \hat{a}^{\dagger} ) / \sqrt{2}$ and $\hat Y_a = ( \hat{a} - \hat{a}^{\dagger} ) / i \sqrt{2}$ for the optical mode, and $\hat X_b = ( \hat{b} + \hat{b}^{\dagger} ) / \sqrt{2}$ and $\hat Y_b = ( \hat{b} - \hat{b}^{\dagger} ) / i \sqrt{2}$ for the microwave mode. The corresponding input noise quaduatures are $\hat X_m^{in} = ( \hat{m}^{in} + \hat{m}^{in,\dagger} ) / \sqrt{2}$, $\hat Y_m^{in} = ( \hat{m}^{in} - \hat{m}^{in,\dagger} ) / i \sqrt{2}$, $\hat X_a^{in} = ( \hat{a}^{in} + \hat{a}^{in,\dagger} ) / \sqrt{2}$, $\hat Y_a^{in} = ( \hat{a}^{in} - \hat{a}^{in,\dagger} ) / i \sqrt{2}$, $\hat X_b^{in} = ( \hat{b}^{in} + \hat{b}^{in,\dagger} ) / \sqrt{2}$, and $\hat Y_b^{in} = ( \hat{b}^{in} - \hat{b}^{in,\dagger} ) / i \sqrt{2}$. In this way, Eq.(5) become
\begin{equation}
\begin{split}
&\hat X_m = -\kappa_m \hat X_m + \Delta_m \hat Y_m - G_{ma} \hat Y_a + g_{mb} \hat Y_b + \sqrt {2{\kappa _m}} \hat X_m^{in},\\
&\hat Y_m = -\Delta_m \hat X_m - \kappa_m \hat Y_m - G_{ma} \hat X_a - g_{mb} \hat X_b + \sqrt {2{\kappa _m}} \hat Y_m^{in},\\
&\hat X_a = - G_{ma} \hat Y_m - \kappa_a \hat X_a + \Delta_a \hat Y_a + \sqrt {2{\kappa _a}} \hat X_a^{in},\\
&\hat Y_a = -G_{ma} \hat X_m - \Delta_a \hat X_a - \kappa_a \hat Y_a + \sqrt {2{\kappa _a}} \hat Y_a^{in},\\
&\hat X_b = g_{mb} \hat Y_m - \kappa_b \hat X_b + \Delta_b \hat Y_b + \sqrt {2{\kappa _b}} \hat X_b^{in},\\
&\hat Y_b = - g_{mb} \hat X_m - \Delta_b \hat X_b - \kappa_b \hat Y_b + \sqrt {2{\kappa _b}} \hat Y_b^{in}. 
\end{split}
\end{equation}
These equations can be rewritten as the following matrix form
\begin{equation}
\begin{split}
\dot u(t) = Au(t) + n(t),
\end{split}
\end{equation}
where $u(t) = [\hat X_m(t),\hat Y_m(t),\hat X_a(t),\hat Y_a(t),\hat X_b(t),\hat Y_b(t){]^T}$, $n(t) = [\sqrt {2{\kappa _m}} \hat X_m^{in},\sqrt {2{\kappa _m}} \hat Y_m^{in},\sqrt {2{\kappa _a}} \hat X_a^{in},\sqrt {2{\kappa _a}} \hat Y_a^{in},\\\sqrt {2{\kappa _b}} \hat X_b^{in},\sqrt {2{\kappa _b}} \hat Y_b^{in}{]^T}$ (The notation $T$ means matrix transport), and the drift matrix
\begin{equation}
\begin{split}
A=\begin{pmatrix}
-\kappa_m & \Delta_m & 0 & -G_{ma} & 0 & g_{mb}\\ 
-\Delta_m & -\kappa_m & -G_{ma} & 0 & -g_{mb} & 0\\ 
0 & -G_{ma} & -\kappa_a & \Delta_a & 0 & 0\\ 
-G_{ma} & 0 & -\Delta_a & -\kappa_a & 0 & 0\\ 
0 & g_{mb} & 0 & 0 & -\kappa_b & \Delta_b\\ 
-g_{mb} & 0 & 0 & 0 & -\Delta_b & -\kappa_b
\end{pmatrix}.
\end{split}
\end{equation}
Owing to the linearity of QLEs and Gaussian nature of the quantum noise, the steady state of the system is a three-mode Gaussian state, fully characterized by covariance matrix $V$ defined as $V_{ij}=\left \langle u_i u_j + u_j u_i \right \rangle / 2$, which can be obtained by solving the Lyapunov equation \citep{ref41}
\begin{equation}
\begin{split}
AV + V{A^T} =  - D.
\end{split}
\end{equation}
in which the diffusion matrix is defined as $D = Diag[\kappa _m (2N(\omega _m) + 1),\kappa _m (2N(\omega _m) + 1),\kappa _a, \kappa _a, \kappa _b (2N(\omega _b) + 1),\kappa _b (2N(\omega _b) + 1)]$. The stability of the system is checked by the Routh-Hurwitz critetion \citep{ref42, ref43}. That is, if all eigenvalues of the drift matrix $A$ hold negative real parts, the system will reach a steady state. However, the exact expression is too cumbersome, so it is omitted here, and all parameters used in this work will fulfill the Routh-Hurwitz critetion unless specifically stated. The entanglement, quatified by logarithmic negativity in this work, between light and microwave can be derived from the matrix
\begin{equation}
\begin{split}
V_{ab}=\begin{pmatrix}
V_1 & V_3\\ 
V_3^{T} & V_2
\end{pmatrix},
\end{split}
\end{equation}
obtained by tracing out rows and columns correlated with magnon in $V$, so does for other two bipartite subsystems, then the entanglement is given by \citep{ref44, ref45}
\begin{equation}
\begin{split}
E_{N} = \max[0,-\ln 2\eta ^{-}],
\end{split}
\end{equation}
where $\eta ^{-}\equiv \sqrt{\Sigma V_{ab}-\sqrt{(\Sigma V_{ab})^{2}-4\det V_{ab}}} / \sqrt{2}$ and $\Sigma V_{ab}\equiv \det V_1 + \det V_2 - 2 \det V_3$.

\section{Results}
\begin{figure}
	\centering
	{
		\label{fig:Fig2(a).png}
		\includegraphics[width=1\linewidth]{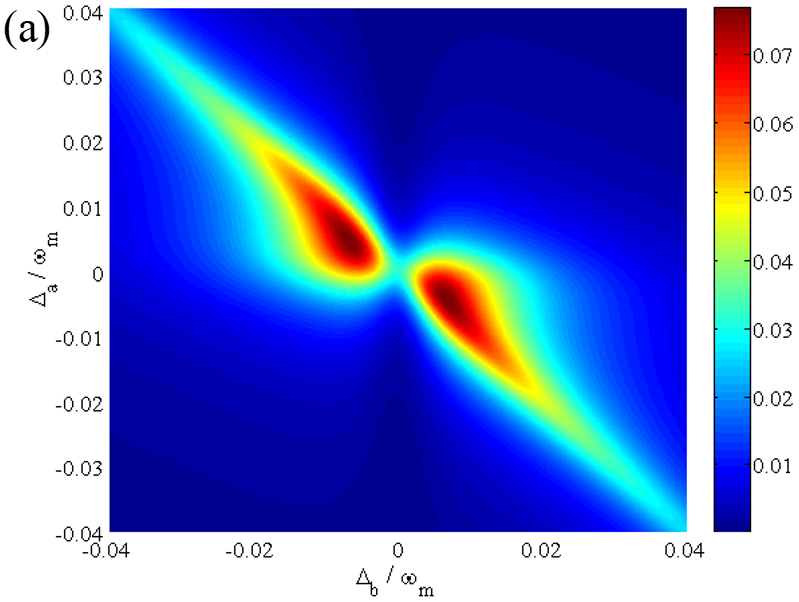}}
	\hspace{1in}
	{
		\label{fig:Fig2(b).png}
		\includegraphics[width=1\linewidth]{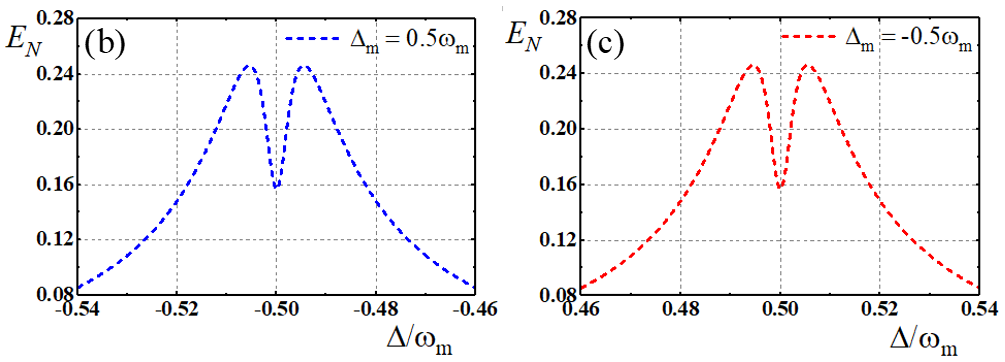}}
	\caption{(a) Density plot of L-M entanglement versus detunings $\Delta_a$ and $\Delta_b$. The magnon detuning $\Delta_m$ = 0, and the quality factor of YIG sphere $Q$ = 2$\times 10 ^{7}$. (b) L-M entanglement versus detuning $\Delta_m$, where $\Delta_a$ = $-\Delta_b \equiv \Delta$ and $Q$ = 5$\times 10 ^{7}$. The common parameters are: electromagnonical coupling $g_{mb} / 2 \pi$ = 6.8 MHz, temperature $T$ = 10 mK.}
	\label{fig:Fig2.png}
\end{figure}

In this section, we present the results of entanglement between light and microwave, and study the entanglement properties of bipartite subsystems in hybrid system. The parameters used in simulation for magnon mode: damping rates $\kappa_m / 2 \pi$ = 1 MHz, external bias magnetic field $B_0$ = 100 mT, and the radius of the YIG sphere $r$ = 125 $\mu$m; for optical TE mode $\hat{a}_1$: pump power $P_p$ = 15 mW, damping rates $\kappa_{a1}$ = $\kappa_a$, pump wavelength $\lambda_p = \frac{2 \pi c}{\omega_p}$ = 1550 nm (so does for optical modes $\hat{a}_2$), pump angular frequency $\omega_{p}$, and its intra-cavity photon number $ \bar{n}_p = \alpha ^2 = \frac{4}{\kappa_{a1}}\frac{P_p}{\hbar \omega_p}$ \citep{ref23}; for microwave part: resonance frequency $\omega_b$ = 9 GHz, damping rates $\kappa_b / 2 \pi$ = 1 MHz; other parameters related are below each figure.

The primary task of studying the characteristics of entanglement properties in such hybrid system is to find the optimal detunings $\Delta_a$, $\Delta_b$, and $\Delta_m$, in other word, to discover the ideal effective interactions among these modes which can generate wanted entanglement between them. In Fig.2(a), show the stationary L-M entanglement as a function of the optical detuning $\Delta_a$ and microwave detuning $\Delta_b$, by setting the magnon detuning fixed at $\Delta_m$ = 0. We find that the largest entanglement is obtained near the case $\Delta_a = -\Delta_b$, which corresponds to the related works that two subsystems bridged by a mediated system can get optimal entanglement when their detunings are opposite \citep{ref17, ref19}. In this case, we shall set these two detunings to be opposite in the following, and for simplicity, we use one symbol to represent them: $\Delta_a$ = $-\Delta_b \equiv \Delta$. In Fig.2(b), we show the relationship between L-M entanglement and magnon detuning $\Delta_m$. We find that the different values of magnon detuning only change the resonance point of the L-M entanglement and do not affect the profile of the entanglement curve. Therefore, without loss of generality, we let the $\Delta_m$ = 0 in the latter.
\begin{figure}
	\centering
	\includegraphics[width=1\linewidth]{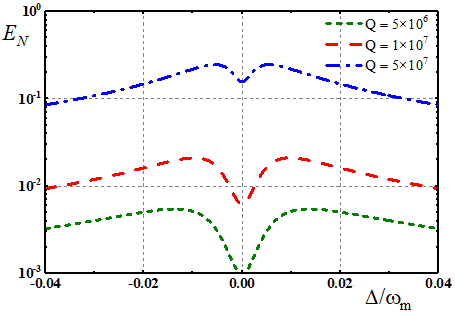}
	\caption{Plot of L-M entanglement versus $\Delta$ with different quality factor $Q$. Green line: $Q$ = 5$\times 10^6$, red line: $Q$ = 1$\times 10^7$, blue line: $Q$ = 5$\times 10^7$. The other parameter are the same as Fig.2.}
	\label{fig:Fig3.png}
\end{figure}

We further study the impact of quality factor on the L-M entanglement in Fig. 3. 
As mentioned above, the pump wavelength are the same for both mode $\hat{a}_1$ and $\hat{a}_2$, so these two optical modes share the same quality factor of YIG sphere. The quality factor $Q$ affects intra-cavity photon number $\bar{n}_p$ of the first optical mode, thus influence the effective optomagnonical coupling $G_{ma}$ between the magnon mode and the second optical mode, and also modulates the damping rate $\kappa_a$, which in general plays an essential role in L-M entanglement. We can find that, as the quality factor increases, the L-M entanglement gains a lot because of the enhancement of optomagnonical coupling, but we cannot generate extremely large entanglement by rudely increasing $Q$ due to the limitation of the system's stability. However, such high quality factor of YIG sphere in telecom band is challenging in current experimental implementations, because of the low telecom photon absorption of YIG material and surface roughness of YIG sphere \citep{ref24}. The potentially feasible solutions are that use other magnetic materials with larger Verdet constant, like $\chemfig{CrBr_3}$ with Verdet constant 8700 rads/cm for the light at 500 nm at 1.5 K environment \citep{ref39, ref46}, to replace the YIG material for enhancing the optomagnonical coupling, or resort to advanced micro- and nanofabrication technology to reduce the surface roughness, thus enhancing the quality factor of YIG sphere. Furthermore, under some certain conditions, the interaction between optical photon and magnon can reach the strong coupling regime \citep{ref32, ref33, ref47, ref48, ref49}, which can be utilized to generate significant L-M entanglement based on our model.
\begin{figure}
	\centering
	\includegraphics[width=1\linewidth]{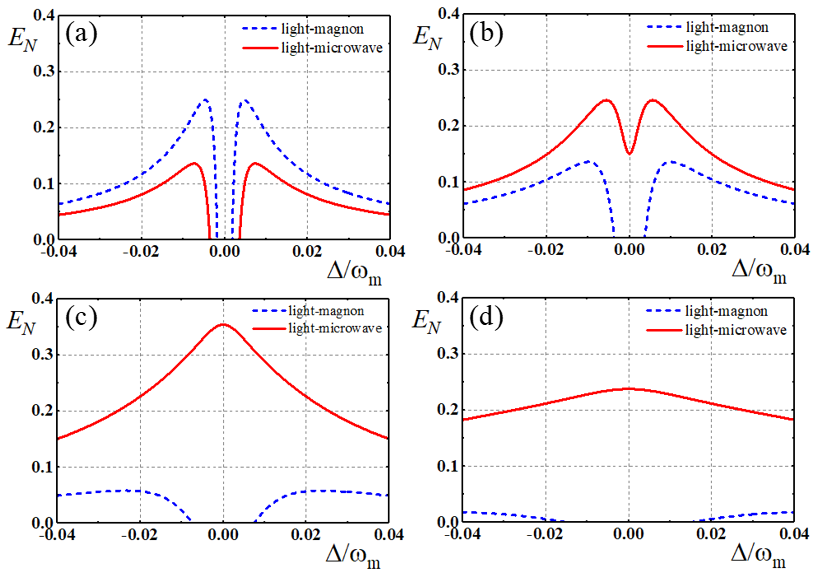}
	\caption{The plot of light-magnon entanglement for blue-dotted line and L-M entanglement for red-full line. (a) $g_{mb} \equiv g_{base} = 2 \pi$ $\times$ 3.4 MHz, (b) $g_{mb} = 2 g_{base}$, (c) $g_{mb} = 4 g_{base}$, (d) $g_{mb} = 8 g_{base}$. The quality factor $Q$ = 5$\times 10 ^{7}$, and the other parameter are the same with Fig.2.}
	\label{fig:Fig4.png}
\end{figure}
\begin{figure}
	\centering
	\includegraphics[width=1\linewidth]{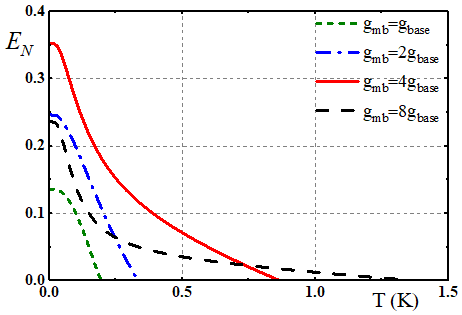}
	\caption{The plot of L-M entanglement with various electromagnonical coupling. Parameters: $g_{base} = 2 \pi$ $\times$ 3.4 MHz, green line $g_{mb} = g_{base}$, blue line $g_{mb} = 2 g_{base}$, red line $g_{mb} = 4 g_{base}$ and black line $g_{mb} = 8 g_{base}$. The other parameters are identical with Fig.4.}
	\label{fig:Fig5.png}
\end{figure}

We turn to investigate the entanglement properties of three bipartite subsystems and analyze the influence from electromagnonical coupling $g_{mb}$ on the entanglement. From Fig.4, we can find that the entanglement only exists between light-magnon and light-microwave, and there is no entanglement between the microwave mode and the magnon mode. This can be explained by the type of interaction among them: the optomagnonical interaction is a two-mode squeezing term, which can generate the optomagnonical entanglement between optical and magnon mode, while the electromagnonical interaction is a beam-splitter term, mainly to transfer the non-classical correlation from one part to the other part under our parameter regime. The entanglement generated by optomagnonical interaction between optical and magnon mode can be partly transferred, thus building the entanglement between light and microwave. We also find that, as $g_{mb}$ increases, the optomagnonical entanglement decreases and the L-M entanglement gets larger. But below a certain value of $g_{mb}$, the L-M entanglement disappears near zero detuning as shown in Fig.4(a), because the system can not fulfill the Routh-Hurwitz criterion, i.e., it will not reach a steady state at that small detuning interval; over a certain value of $g_{mb}$, the maximum L-M entanglement declined, in part due to the overcoupling between microwave and magnon modes, which guides us to find the optimum $g_{mb}$ for the optimum L-M entanglement.

Fig.5 shows the plot of L-M entanglement as a function of temperature with various electromagnonical coupling, and it shows that the entanglement is robust against thermal environment under our parameter regime. Over a certain value of $g_{mb}$, the L-M entanglement will decline by increasing the $g_{mb}$ as discussed above, but the robustness facing higher-temperature environment is always enhanced by increasing the electromagnonical coupling. It is shown that the L-M entanglement still persist above 1.2 K with experimentally feasible $g_{mb}$.

Finally, let us briefly discuss how to detect the L-M entanglement. As discussed above, the entanglement is calculated from the covariance matrix $V$, so we can obtain the wanted results by measuring the corresponding $V$ \citep{ref41, ref50, ref51, ref52}. The state of magnon in YIG sphere can be acquired by adding an external microwave probe field through the YIG sphere and homodyning its output, while the microwave and optical field quadratures can be directly measured by homodyning their output fields too.

\section{Conclusion}
We have proposed a scheme to generate stationary and robust entanglement between light and microwave modes mediated by magnon modes in YIG sphere, which is considered as a new promising platform for building hybrid quantum systems. Our results show some similarities with one previous work that uses the nanomechanical resonator as a bridge to generate the entanglement between light and microwave \citep{ref16}, which indicates that one can also use our model to carry out some quantum tasks, like reversible quantum interface between light and microwave photons \citep{ref17} and microwave quantum illumination \citep{ref13}. Compared with other works using logarithmic negativity as the entanglement measure that is related to the L-M entanglement and magnon-based systems \citep{ref16, ref29, ref30, ref53}, the entanglement in our results shows the same order with them. The value of the entanglement between light and microwave via the optomechanical system in Ref.\citep{ref16} is about 0.2; the value of magnon-magnon entanglement and microwave-magnon entanglement in Ref.\citep{ref29, ref30} is about 0.15; the value of the entanglement between two microwave modes via their common interaction with magnon modes in the electromagnonical system described in Ref.\citep{ref53} is about 0.15. The simulation of light-microwave entanglement value in our work could reach above 0.2, which shows the almost identical quality or a little higher compared to these works.

Almost all simulation parameters are reachable in state-of-the-art experiments except the high quality factor of YIG sphere for optical modes, which is aimed for strong optomagnonical coupling, and the present experimentally feasible quality factor is $\sim 10^6$ \citep{ref24}. But, we also mention some potential solutions in the Results part to enhance the quality factor of YIG sphere and reach the strong-coupling regime in optomagnonics, such as replacing the YIG material with $\chemfig{CrBr_3}$ possessing larger Verdet constant, reducing the surface roughness of YIG sphere by advanced fabrication technology, and others \citep{ref32, ref33, ref46, ref47, ref48, ref49}. What is more, the frequency mismatch in experiments between the TE, TM WGMs and magnon frequency may also limit the L-M entanglement, however, this problem could be tackled by properly designing the geometries of the WGM resonator. It is to be expected that these difficulties will eventually be overcome, once realized, our work may open a new perspective to realize the quantum interface between various physical objects, such as photon, mechanical membrane, NV center, quantum dot, atom and so on, leading to more advanced and comprehensive hybrid quantum systems to fulfill the increasing requests in quantum technologies.

\bibliography{Magnon.bib}
\end{document}